\documentclass[a4paper,fleqn,usenatbib,useAMS]{jaa}

\usepackage{graphicx}	
\usepackage{amsmath}
\usepackage{amssymb}	
\usepackage{multicol}        
\usepackage{bm}		
\usepackage{pdflscape}	
\usepackage{psfrag}
\def\xb{\bar{x}_{\rm HI}}

\def\HI{{\rm HI}}

\def\prd{PhReD}

\def\mnras{MNRAS}

\def\aap{AAP}
\def\apj{Ap.J}

\def\HI{{\rm HI \,}}

\def\V{{\mathcal V}}

\def\del{\partial}

\def\x{\mathbf{x}}

\def\lsim{~\rlap{$<$}{\lower 1.0ex\hbox{$\sim$}}}

\def\gsim{~\rlap{$>$}{\lower 1.0ex\hbox{$\sim$}}}

\usepackage[T1]{fontenc} \usepackage{ae,aecompl}

\pubyear{2016}
\begin{document}
\title[HI - OWFA]{Simulating the z = 3.35 HI 21-cm visibility signal
  for the Ooty Wide Field Array(OWFA)} \author[Suman Chatterjee,
  S. Bharadwaj and V. R. Marthi] {Suman
  Chatterjee$^{1,2}$\thanks{Email:suman05@phy.iitkgp.ernet.in}, Somnath
  Bharadwaj$^{1,2}$\thanks{Email:somnath@phy.iitkgp.ernet.in} and
  \newauthor Visweshwar Ram Marthi
  $^{3}$\thanks{Email:viswesh@ncra.tifr.res.in}\\  $^{1}$ Department
  of Physics,  IIT Kharagpur, 721302, India \\ $^{2}$ Centre for
  Theoretical Studies, IIT Kharagpur, 721302 , India \\ $^{3}$
  National Centre for Radio Astrophysics, Tata Institute of
  Fundamental Research,\\ Post Bag 3, Ganeshkhind, Pune - 411 007,
  India} \date {} \maketitle
%
%
\begin{abstract}
The upcoming Ooty Wide Field Array (OWFA) will operate  at  $326.5 \,
{\rm MHz}$  which corresponds  to  the redshifted  21-cm signal  from
neutral hydrogen (HI) at z = 3.35.  We present two different
prescriptions to  simulate this signal and calculate the visibilities
expected in   radio-interferometric  observations with OWFA.  In the
first method we use an input model for the expected 21-cm   power
spectrum  to directly simulate different random realizations of  the
brightness temperature fluctuations and calculate the visibilities.
This method, which models the HI signal entirely as a diffuse
radiation,  is completely oblivious to the discrete nature of the
astrophysical sources  which host the HI.  While each  discrete
source  subtends an angle that is much smaller than the  angular
resolution of  OWFA, the velocity structure of the HI inside the
individual sources  is well within reach of OWFA's frequency
resolution and this is expected  to  have an impact on the observed
HI signal. The second prescription  is based on  cosmological N-body
simulations. Here we identify each simulation  particle with a  source
that hosts the HI, and we have the freedom to implement any desired
line profile for the HI emission from the individual sources.
Implementing a simple model for the line profile, we have generated
several  random realizations of the complex visibilities. Correlations
between the visibilities measured at different baselines and channels
provides an unique method to quantify the statistical properties of
the \HI signal. We have used this to quantify the results of our
simulations, and explore the relation between the expected visibility
correlations  and the underlying HI power spectrum. 
\end{abstract}

%
%
%
%
%
\section{Introduction}
The redshifted neutral hydrogen (HI) 21-cm radiation  is present as a
background radiation in all low-frequency radio observations below
$1420 \, {\rm MHz}$. A statistical detection of the random
fluctuations in this background radiation  present an unique technique
for probing the large scale  structures in the universe during the
post-reionization era $(z \le 6)$  (Bharadwaj, Nath \& Sethi 2001;
Bharadwaj \& Sethi 2001).  Observations of high redshift quasars  show
that the diffuse  inter-galactic medium (IGM) is highly ionized at
redshifts $z < 6$, and the  bulk of the HI in the post-reionization
era  resides in discrete, dense HI clouds with column densities
greater than $2 \times 10^{20} \, {\rm atoms}\,{\rm cm}^{-2}$ (Wolfe et
al. 1995; Lanzetta, Wolfe \& Turnshek 1995; Storrie-Lombardi, McMahon \& Irwin 1996;
 P{\'e}roux et al. 2003).  Several measurements of $\Omega_{\HI}(z)$ and
  $\Omega_{\rm gas}\left( z \right)$ have been carried out both at low and
   high redshifts. At low redshifts ($z \sim 1$ and lower) we have measurements 
   of $\Omega_{\HI}(z)$ from HI galaxy surveys (Zwaan et al. 2005;
Martin et al. 2010; Delhaize et al. 2013) and  HI stacking (Lah et al. 2007; Rhee et al. 2013, 2016).
Observations of the Damped Ly$-{\alpha}$ (DLA) systems
provide us measurements of $\Omega_{\rm gas}\left( z \right)$  for both
low redshifts ($z \sim 1$, Rao, Turnshek \& Nestor  2006; Meiring et al. 2011) and
high redshifts($2< z < 6$, Prochaska \& Wolfe 2009; Noterdaeme et
al. 2012). These observations tell us that the HI content of the
universe remains almost constant $\Omega_{\rm gas}\left( z \right) \sim
10^{-3}$ over the entire redshift range $z<6$ (Lanzetta et al. 1995;
Storrie-Lombardi et al. 1996; Rao \& Turnshek 2000; P{\'e}roux et
al. 2003).  Unlike traditional galaxy redshift surveys which  need to
identify individual galaxies in order to map out the large scale
structures in the universe, the proposed 21-cm surveys do not need
to identify the  individual HI sources. The collective redshifted
21-cm radiation from  the  individual HI  clouds appears  as  a
diffuse background  radiation and the source clustering is imprinted
as the fluctuations in this background  radiation.

   Future redshifted 21-cm observations are currently perceived to be a very promising
probe of the formation and evolution of large-scale structures  in the
post-reionization era (Bharadwaj \& Pandey 2003; Bharadwaj \& Ali 2005;
Wyithe \& Loeb  2008).  Such observations hold the potential of
measuring the Baryon Acoustic Oscillation (BAO) which  is a very
sensitive  probe of dark energy (Wyithe, Loeb \& Geil 2008; Chang et
al.  2008; Seo, Dodelson \& Marriner 2010; Masui, McDonald \& Pen   2010).  Further, a
measurement of just the 21-cm power spectrum can also be used to
constrain cosmological parameters (Visbal, Loeb \& Wyithe 2009;
Bharadwaj, Sethi \& Saini 2009). 

The Giant Meterwave Radio Telescope (GMRT, Swarup et al. 1991) is
sensitive to the cosmological HI signal from a range of redshifts in
the post-reionization era  (Bharadwaj \& Pandey 2003, Bharadwaj \& Ali
2005) and Ghosh et al. (2011a,b) have carried out preliminary
observations towards detecting this signal from $z = 1.32$.  The
upgraded uGMRT is expected  to have a larger bandwidth for which the
prospects of a detection are presented in  Chatterjee et al. (2017).
The Canadian Hydrogen Intensity Mapping Experiment (CHIME, Newburgh et
al. 2014 , Bandura et al. 2014)  aims to measure the BAO in the
redshift range $0.8 - 2.5$.   The future Tianlai (Chen 2012, 2015) and
SKA1-MID   (Bull et al. 2015) also aim to measure the redshifted HI
21-cm signal  from the post-reionization era. 

The Ooty Radio Telescope (ORT; Swarup et al. 1971) is currently  being
upgraded (Prasad \& Subrahmanya 2011, Subrahmanya et al. 2017b)    to
a linear radio-interferometric array, the Ooty Wide Field Array (OWFA;
Subrahmanya, Manoharan \& Chengalur 2017a). The OWFA  operates at a
nominal frequency of $\nu_o=326.5 \, {\rm  MHz}$ which corresponds to
the redshifted 21-cm signal from a redshift $z= 3.35$.  Measuring the
redshifted HI 21-cm power spectrum is one of the major science goals
of OWFA (Subrahmanya et al. 2017a).  Theoretical estimates (Bharadwaj,
Sarkar \& Ali 2015) predict that it should be possible to measure the amplitude of the
21-cm power spectrum  with $150 \, {\rm hrs}$ of observations using
OWFA. A more recent study (Sarkar, Bharadwaj \& Ali 2017) indicate
that it should be possible to measure the 21-cm power spectrum in
several different $k$ bins in the range $0.05 - 0.3 \, {\rm Mpc}^{-1}$
with $1,000 \, {\rm hrs}$ of observations.  

The complex visibilities are the primary quantities measured by any
radio-interferometric array like OWFA.  It is  important and
interesting to directly quantify the  HI signal in terms of the
expected contribution to the measured visibilities.  Ali \& Bharadwaj
(2014) present detailed predictions for the correlations between the
visibilities measured at different baselines and frequency channels
expected from the statistical HI signal, various astrophysical
foregrounds and the system noise. Similar predictions for the HI
signal incorporating  non-linear effects  are also presented in a more recent
paper (Gehlot \& Bagla 2017).  These theoretical estimates   predict that the 
visibilities 
measured at OWFA will be dominated by astrophysical foregrounds which
are expected to be several orders of magnitude larger than the HI
signal. Foreground removal (e.g. Ghosh et al. 2011b) is therefore an
important issue for detecting the redshifted 21-cm signal.  The
astrophysical foregrounds are all expected to have a smooth frequency
dependence in contrast to the HI signal, and most foreground removal
techniques rely on this to distinguish between the foregrounds and the
HI signal. However, the chromatic response of the telescope
introduces frequency structures in the  foregrounds, and it is
necessary to model these and account for these in any foreground
removal technique.  Here, simulations play a crucial role in modelling
the impact of various complicated effects like the telescope's
chromatic response, calibration (Marthi \& Chengalur 2014) etc.
Simulations which incorporate the expected foregrounds, HI signal and
also various instrumental and post-processing effects are essential
for testing and validating any foreground removal and power spectrum
estimation technique. 

Marthi (2017) has developed a programmable emulator for OWFA, this
provides a platform for simulating the visibilities that will be
measured at OWFA.   The foreground modelling and predictions are
presented in Marthi et al. (2017). In this paper we focus on modelling
the HI signal contribution to the visibilities that will be measured
at OWFA. 

Bagla, Khandai \& Datta  (2010) have proposed  a semi-numerical
technique to simulate the redshifted HI signal. This uses N-body
simulations to predict the locations and masses of dark matter halos,
and relies on a prescription to populate the halos with HI. The same
technique has been utilized in subsequent work by  Khandai et
al.(2011), Guha Sarkar et al.(2012),  Villaescusa-Navarro et al. (2014)  and
Sarkar, Bharadwaj \& Anathpindika (2016). It may be noted that this technique requires the
N-body simulation to have a rather high spatial resolution $\sim 0.1
\, {\rm Mpc}$ (comoving) in order to resolve the halos which host the
HI. In comparison, an angular extent of  $1^{\circ}$  and a
frequency spread  of  $1 \,{\rm MHz}$ respectively corresponds to
comoving distances of $\sim 113 \, {\rm  Mpc}$ in the transverse
direction and   $\sim 11.5 \, {\rm Mpc}$ along the line of sight at
$z=3.35$.   Considering the fact that OWFA's field of view (FoV)
subtends several degrees on the sky and the bandwidth covers several
tens of ${\rm MHz}$ in frequency, it is computationally prohibitive to
implement this technique to directly simulate the signal and
calculate the visibilities expected at a Wide Field Array  like OWFA. 

In this paper we present two different methods  to  simulate the HI
signal and calculate the visibilities expected in   observations with
OWFA.  In the first method we use the results of the semi-numerical
simulations (e.g. Bagla, Khandai \& Datta 2010) to construct  an input
model for the expected 21-cm   power spectrum. This is used     to
directly simulate different random realizations of  the   brightness
temperature fluctuations and calculate the visibilities.The HI signal
here is  assumed to be a Gaussian random field.  This method  models
the HI signal entirely as a diffuse  radiation, and   is completely
oblivious to the discrete nature of the clouds  which host the HI. 

As noted above, the discrete clouds  which  host the HI each   subtends an angle ($
\sim 1^{''}$ or smaller)  that is much smaller than the  angular
resolution ($0.1^{\circ}$) of  OWFA.  The velocity structure $(\sim
100 \, {\rm km/s})$ of the HI inside the individual clouds, is
however  well within reach of OWFA's frequency resolution  ($\sim 100
\, {\rm km/s}$ or smaller), and this is expected  to  have an impact
on the observed HI signal. We have also used a second prescription, which   is based
on earlier work by Bharadwaj \& Srikant (2004), uses a  cosmological
N-body simulation to generate the matter distribution at the desired
redshift. Here we identify each simulation  particle with a  source
that hosts the HI and use this to calculate the expected
visibilities. This technique provides us with  the freedom to
implement any desired line profile for the HI emission from the
individual sources. The simulations presented in this paper
are preliminary, thermal noise and complicated systematics have not 
 been included here.

A brief outline of the paper follows. In section 2 we present the
model 21-cm power spectrum that was used to simulate the HI signal,
and Section 3 provides a brief overview of OWFA.  The first (direct)
simulation technique is  presented in Section 4, which also contains
the results from this approach.   The second (particle based)
technique, along with the results, are presented in Section 5.  The
results from both the simulation techniques are discussed  and
summarized in Section 6.

We have used the fitting formula of Eisenstein \& Hu(1998) for the
$\Lambda$-CDM transfer function to generate the initial, linear
matter power spectrum. The  cosmological parameter values used are as
given in Ade et al.(2014):$\Omega_m = 0.318$, $\Omega_b \, h^2 = 0.022$,
 $\Omega_{\lambda} = 0.682$, $n_s = 0.961$, $\sigma_8 = 0.834$, $h = 0.67$.
\section{The 21-cm Power Spectrum}
We model the 3D spatial  power spectrum $P_T(\mathbf{k})$ of the
redshifted 21-cm brightness temperature fluctuations  at $ z = 3.35$
as   
\begin{equation}
P_T\left( \mathbf{k}\right)  = \bar{T}^2 \;\bar{x}^2_{\HI} \;
P_{\HI} \left( \mathbf{k}\right)
\label{eq:temp_pow}
\end{equation}
where,
\begin{equation}
\bar{T}\left( z\right) = 4.0 \, mK \; \left( 1+ z \right)^2 \;
\left(\frac{\Omega_b h^2}{0.02} \right) \; \left(\frac{0.7}{h} \right)
\;  \left( \frac{H_0}{H\left(z \right) } \right) 
\label{eq:b_temp}
\end{equation}
and $P_{\HI} \left( \mathbf{k}\right)$ is the power spectrum of the HI
density fluctuations, all other symbols have their usual meaning
(Bharadwaj \& Ali 2005). As mentioned earlier, DLA observations
indicate $\Omega_{\rm gas} \simeq 10^{-3}$. This implies that the mass averaged
neutral hydrogen fraction has a value $\bar{x}_{\HI} = 2.02 \times
10^{-2}$  which we have used in our work.
Assuming that the HI gas traces the underlying matter
distribution  with a linear bias $b_{\HI}$, 
we relate $P_{\HI} \left(\mathbf{k}\right)$  to 
the underlying matter power spectrum $P(\mathbf{k})$   as 
\begin{equation}
P_{\HI} \left( \mathbf{k}\right) = b^2_{\HI}\; \left( 1+\beta \mu^2 \right)
^2 \; P(\mathbf{k})
\label{hi_pow}
\end{equation}
where $\mu = k_{\parallel}/k$, and $k_{\parallel}$ is the component of the
comoving wave vector $\mathbf{k}$ along the line of sight. 
 The term $\left( 1+\beta \mu^2 \right)
$ arises due to of the effect of HI peculiar velocities (Bharadwaj, Nath \&
Sethi 2001; Bharadwaj \& Ali 2005),  and $\beta = f(\Omega)/b_{\HI}$
is the linear redshift distortion parameter, where $f(\Omega)$ is the 
dimensionless linear growth rate. 

We currently do not have any observational constraint on $b_{\HI}$ for
high redshifts. However, the results from the different semi-numerical simulations,
and also an interpolated compilation of $b_{\HI}$ values 
(Padmanabhan, Roy Choudhury \& Refregier 2015), indicate that it is
reasonably well justified to assume a constant HI bias $b_{\HI}= 2$
at wave numbers $k \leq 1 \, {\rm Mpc}^{-1}$ at the redshift $z = 3.35$. We have
used this value and $\beta=0.493$ for our entire analysis,  this corresponds to
$f(\Omega) = 0.986$ (Bharadwaj,Sarkar \& Ali 2015). 
\section{A Brief overview of OWFA}
 In this paper we consider the so called ``Phase-I'' of OWFA (e.g. Ali \&
 Bharadwaj 2014, Subrahmanya, Manoharan \& Chengalur 2017a) where  $24$ successive dipoles  are combined to
 form a single antenna which gives rise to a  $40$ element interferometer. Each
 antenna element has a  rectangular aperture $b \times d$ of dimensions $b =
 30 \, {\rm m}$ along E-W and $d = 11.5 \, {\rm m}$ along N-S respectively. 
This gives a field-of-view (FoV)of $1.8^{\circ}$ (E-W) $\times 4.5^{\circ}$ (N-S).
 The 40 antenna elements are equally spaced
in a linear array with a separation of $11.5 \, {\rm m}$ along the N-S direction. The
telescope is expected to have a frequency bandwidth $B_{bw} = 39  {\rm MHz}$. We
assume that the whole bandwidth is divided in $N_c = 312$ frequency
channels of $\Delta \nu_c = 125 \, {\rm KHz}$ channel width. Note that the
values of $B_{bw}$, $N_c$ and $\Delta \nu_c$ assumed here are only representative
values, and the actual values in the final implementation of the telescope may
be somewhat different. 

	
The complex visibilities $\mathcal{V}(\mathbf{U},\nu)$ measured by OWFA will be 
recorded for every independent pair of antennas for 
every available frequency channel. For any pair of antennas, the baseline $\mathbf{U} =
\frac{\mathbf{d}}{\lambda}$ quantifies the antenna separation in units of the
observing wavelength. For observation towards the celestial equator, the
baselines available at OWFA are 
\begin{equation}
\mathbf{U}_n = n \frac{d}{\lambda} \hat{j}	\quad \quad \left( 1
\leq n \leq N_A - 1\right) 
\label{bline}
\end{equation}
where $N_A = 40$ is the total number of antennas for Phase I.  There is considerable
redundancy in the OWFA baseline distribution in that there are 
many different antenna pairs which correspond to the same
baseline. Any baseline $\mathbf{U}_n$ occurs $M_n = N_A -n $ times in
the array. The baseline $\mathbf{U}_n$ corresponding to a fixed antenna
separation also changes with the observing wavelength $\lambda$.

We have used the flat sky approximation for our calculations throughout this
paper.  Here $\bm{\theta}$ is a two dimensional vector in the 
plane of the sky with the origin at the centre of the FoV. The visibilities 
$\mathcal{V}(\mathbf{U},\nu)$ are the Fourier  transform of the product of the
specific intensity 
distribution on the sky $I(\bm{\theta}, \nu)$ and the telescope's primary beam 
pattern $A(\bm{\theta}, \nu)$.  We calculate the visibilities using 
\begin{equation}
\mathcal{V}(\mathbf{U}_n,\nu_a) = \left( \frac{\del B}{\del T}\right)
\int d^2\bm{\theta} A(\bm{\theta}, \nu_a) \;  T(\bm{\theta}, \nu_a)
e^{- 2 \pi i \mathbf{U}_n\cdot \bm{\theta}}
\label{eq:vis1}
\end{equation}
where $B = 2k_B T /\lambda^2$ is the Planck function in the
Raleigh-Jeans limit which is valid at the frequency range of our
interest and $T(\bm{\theta}, \nu)$ is the brightness temperature
distribution on the sky. 
	
The primary beam pattern for the rectangular aperture of the antenna
elements (Ali \& Bharadwaj 2014) can be expressed as
\begin{equation}
A(\bm{\theta}, \nu) = {\rm sinc}^2\left( \frac{\pi b
  \theta_x}{\lambda}\right) {\rm sinc}^2\left( \frac{\pi d
  \theta_y}{\lambda}\right)
\label{eq:pbeam}
\end{equation}
where $\theta_x$ and $\theta_y$ are respectively the $x$ and $y$ 
components of $\bm{\theta}$  in a Cartesian coordinate system that is aligned
with the plane of the  aperture  and 
has the $y$ axis along the N-S direction.  We assume this pattern in 
our simulations, although we note that in practice tapering of the illumination
will cause the OWFA element pattern to have lower side lobes and somewhat
broader
main lobes.

We have decomposed the brightness temperature $T(\bm{\theta}, \nu)$ 
as the sum of an uniform average background $\bar{T}_b(\nu)$ 
and another component $\delta T(\bm{\theta}, \nu)$ which fluctuates with
direction on the sky
\begin{equation}
T(\bm{\theta}, \nu) = \bar{T}_b(\nu)+\delta T(\bm{\theta}, \nu) \,.
\label{eq:fluc}
\end{equation}
In the flat sky approximation, the visibilities measured at OWFA 
do not respond to the uniform background $\bar{T}_b(\nu)$  (Ali \& Bharadwaj
2014), and it is adequate to only consider the fluctuations $\delta
T(\bm{\theta}, \nu)$ in eq. (\ref{eq:vis1}) to calculate the visibilities.

\section{Direct Simulation of Brightness Temperature Fluctuations}
%

In this section  we present simulations where we use an input model for 3D
spatial power spectrum $P_T(\mathbf{k})$ (eq. \ref{eq:temp_pow}) to directly
generate $\delta T(\bm{\theta}, \nu)$(eq. \ref{eq:fluc}) on the sky and
calculate the expected visibilities(eq. \ref{eq:vis1}).  

The $326.5\, {\rm  MHz}$  OWFA observations correspond to HI at 
$z = 3.35$ which is  at a comoving distance of $r=  6.84 \, {\rm 
  Gpc}$. This  sets the conversion scale from angular separation to comoving
distance in the transverse direction. 
We further have  $r' =\mid dr/d\nu \mid= 11.5 \, { \rm Mpc} \, {\rm MHz}^{-1}$
which sets the conversion scale from frequency separation 
to comoving distance in the radial  direction.  The OWFA Phase I FoV 
of $1.8^{\circ}  \times 4.5^{\circ}$ corresponds to $215 \, {\rm Mpc}
\times 550 \, {\rm Mpc}$ in the two transverse directions and the bandwidth of
$B_{bw}=39 \, {\rm MHz}$ corresponds to $449 \, {\rm Mpc}$ in the radial
direction. 

\begin{figure}
\psfrag{RAAAAAAAAAAAA}{Declination (Deg)} \psfrag{DDDDDDECCCC}{Right
  Ascension (Deg)} \psfrag{cccolourrr}{$A(\bm{\theta}, \nu_a)$}
\psfrag{colourrrrr}{$\delta T (\bm{\theta}, \nu_a) {\rm mK} $}
\includegraphics[scale=1.0, angle = 0]{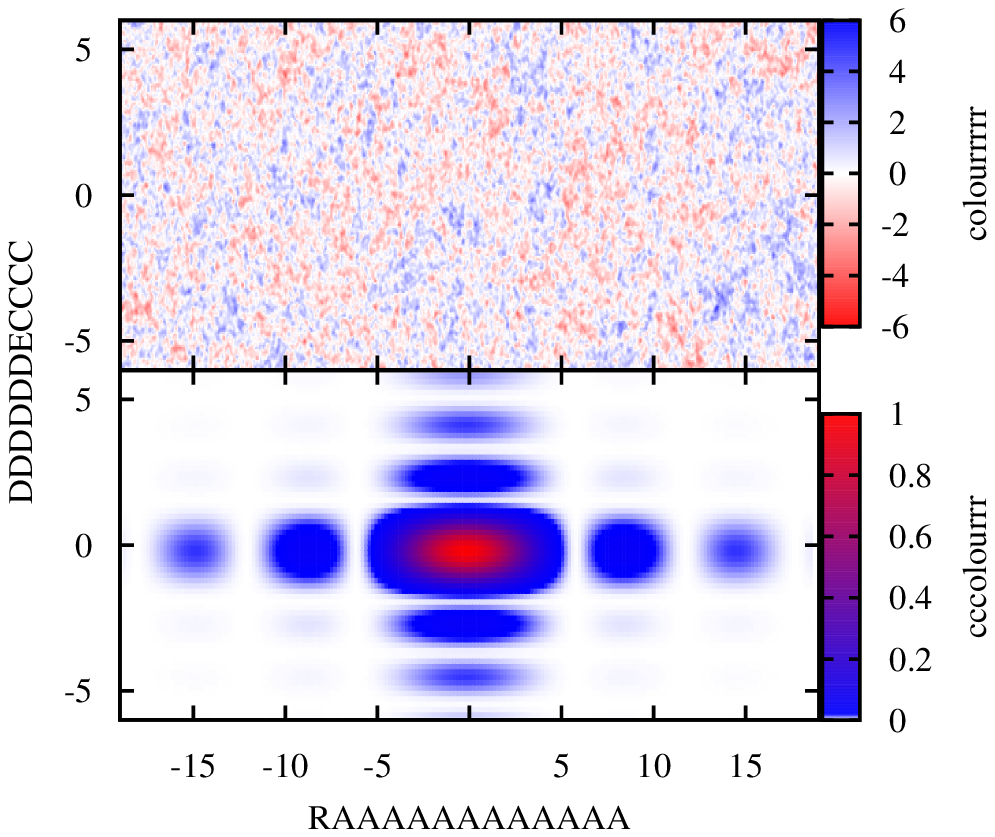}
\caption{ The upper panel shows a single realization of the
simulated redshifted 21-cm brightness temperature fluctuations
$\delta T(\bm{\theta},\nu_a)$ as a function of $\bm{\theta}$ for a fixed 
$\nu_a$ corresponding to the central frequency  channel. 
Note that the angular extent shown here corresponds to the 
region  which was actually cut out from the entire  simulation
volume and used to calculate the OWFA visibilities. 
The lower panel shows a similar plot of the beam pattern
$A(\bm{\theta},\nu_a)$. } 
\label{fig:2}
\end{figure}

The simulations were carried out using a $N^3$ cubic grid of spacing $L$
covering a comoving volume $V$. We have chosen the grid spacing $L = 1.44 \,
{\rm Mpc}$ so that it exactly matches the channel width $L = r'\times
(\Delta \nu_c)$, where the spectral channels are of width $(\Delta \nu_c)=
0.125 \, {\rm MHz}$. We have considered a $N^3 = [2816]^3$ grid which corresponds
to a comoving volume of $[4048 \, {\rm Mpc}]^3$.The considered volume is 
kept few times larger to incorporate the side lobe contributions (Figure \ref{fig:2}),
 that comes from the radiation received in the side lobes of the antenna pattern.

\begin{figure}
\begin{center}
\psfrag{DECCCCCCC}{$\quad P_T(k)\;{\rm mK}^2 \, {\rm Mpc}^3$}
\psfrag{RAAAAAAAAAAAA}{$\quad \quad \quad k \; {\rm Mpc}^{-1}$} 
\includegraphics[scale=0.5,
  angle = 270]{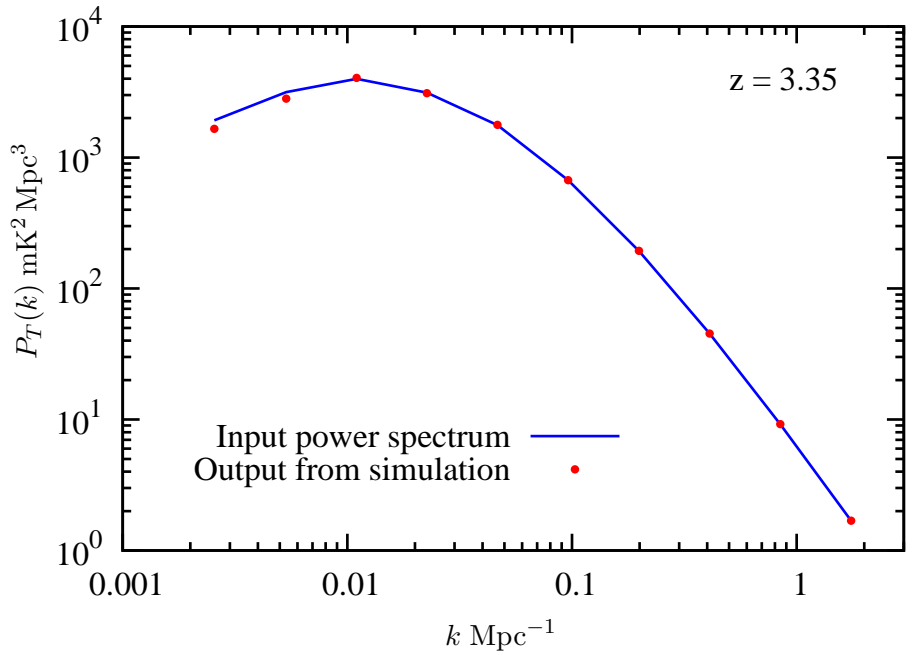}
\end{center}
\caption{The solid blue curve shows the input model  21-cm brightness
  temperature power spectrum $P_T(k)$, and the points show  $P_T(k)$
  calculated from one realization of the simulation.  Both have been averaged
  over $\mu$ to remove the $\mu$ dependence.} 
\label{fig:1}
\end{figure}

We use the input power spectrum $P_T \left( {\bf k} \right)$ to generate the 
Fourier components of the brightness temperature fluctuations 
\begin{equation}
\Delta T \left(\mathbf{k} \right)  = \sqrt{\frac{V P_T \left( {\bf k}
    \right) }{2}} [a \left( \mathbf{k} \right) +\mathit{i} b \left(
  \mathbf{k} \right) ] \,. 
\label{eq:den_k}
\end{equation}
on a grid  of  wave vectors ${\bf k}$ corresponding to the simulation
volume.  Here  $a(\mathbf{k})$ and $b(\mathbf{k})$ are two real valued
independent Gaussian random variable of unit variance which satisfy
\begin{equation}
\langle a(\mathbf{k}) a(\mathbf{k}') \rangle = \langle b(\mathbf{k})
b(\mathbf{k}') \rangle = \delta_{\mathbf{k},\mathbf{k}'} \,,\langle a(\mathbf{k}) b(\mathbf{k}') \rangle = 0 \; \forall \;  \mathbf{k},\mathbf{k}' \, .
\end{equation}
The Fourier transform of $\Delta T(\mathbf{k})$ yields a single
realization of the 21-cm brightness temperature 
fluctuations $\delta T(\mathbf{x})$ on the
simulation grid.  These fluctuations are, by construction, a Gaussian
random field with the input 21-cm power spectrum.  We use different sets of
the random variables $a(\mathbf{k})$ and $b(\mathbf{k})$ to generate different
statistically independent realizations of $\delta T(\mathbf{x})$. Note that
the adopted $P_T(\mathbf{k})$ include the effect of redshift space distortion along the line
of sight. Figure \ref{fig:1}  shows a comparison of the input model power
spectrum with the power spectrum estimated from a single realization  of the
simulation. We find that the simulated power spectrum is in good agreement
with he input model. 

 The simulation volume is aligned with the $z$ axis along the line of
 sight.  This corresponds to a frequency width of 
 $\sim 9 \times 39 \, {\rm MHz}$ along the $z$ axis. We have cut the box
 into 9 equal segments along the line of sight to produce 9
 independent realizations each corresponding to a bandwidth of $39 \, {\rm MHz}$ . 
The grid index, measured from the further boundary and
 increasing towards to observer along the line of sight was directly
 converted to channel number $a = 0, 1, 2, ..., N_c - 1$ whereby $\nu_a= 
307 \, {\rm MHz}+ a L/r^{'}$ .  
The two transverse directions were restricted to 
$1553 \, {\rm Mpc}$ and  $4048 \, {\rm Mpc}$  along the $x$ and $y$ axes
respectively, and these were converted to angles
relative to the center ($\theta_x$, $\theta_y$) = ($x/r$,  $y/r$). 
The extent along the $x$ and $y$ axes were chosen so as to contain
approximately the first three nulls of the beam pattern
$A(\bm{\theta},\nu_a)$  (bottom panel of Figure \ref{fig:2}) along each 
direction. The cut out region of the 
simulation box  procedure provides us with $\delta T(\bm{\theta}, \nu_a)$
 the brightness  temperature fluctuation  on
 the sky (upper panel of Figure \ref{fig:2}) at different frequency  channels
 $\nu_a$.

We calculate the visibilities using  a discretized version of
eq. (\ref{eq:vis1})  whereby 
\begin{equation}
\mathcal{V}(\mathbf{U}_n,\nu_a) = \left( \frac{\del B}{\del T}\right)
(\Delta \theta)^2 \sum_{g}  A(\bm{\theta}_{g},
\nu_a)  \delta T(\bm{\theta}_{g}, \nu_a) e^{- 2 \pi i
  {U}_n \theta_{yg}}
\label{vis2}
\end{equation}
where $\Delta \theta$ is the angular resolution of the simulation grid, the
sum  is over all the grid points (labelled using $g$) 
corresponding to the fixed frequency channel $\nu_a$,  and $\theta_{yg}$ is the
$y$ component of the vector   $\bm{\theta}_g$ corresponding to the 
grid point $g$. 

The baseline and frequency channel configuration of OWFA Phase I probes
 the $k_{\perp}$ range $k_{\perp_{min}} = 1.1 \times 10^{-2} \, { \rm
  Mpc}^{-1}$ to  $k_{\perp_{max}} = 4.8 \times 10^{-1} \, { \rm Mpc}^{-1}$, 
and the  $k_{\parallel}$ range $k_{\parallel_{min}} = 1.4 \times 10^{-2} \, { \rm Mpc}^{-1}$ to
$k_{\parallel_{max}} = 2.73 \, { \rm Mpc}^{-1}$ (Bharadwaj et al. 2015).
For our simulations  we have the corresponding values 
$k_{\perp_{min}} = 1.4 \times 10^{-3} \, { \rm Mpc}^{-1}$, $k_{\perp_{max}} =
4.37 \, { \rm Mpc}^{-1}$, $k_{\parallel_{min}} = 1.4 \times 10^{-2} \, { \rm
  Mpc}^{-1}$ and  $k_{\parallel_{max}} = 2.73 \, { \rm Mpc}^{-1}$. As can be seen
our simulations adequately cover the entire $k$ range that will be probed by
OWFA Phase I.

\subsection{Validating the simulation}
The redshifted 21-cm brightness temperature fluctuations are, by construction, 
a Gaussian random field. The visibilities obtained from the simulations also
are random,  and each realization of the simulation will yield a different set
of values for the visibilities. It is necessary to consider the statistical
properties of the simulated visibilities in order to validate the
simulations. In this paper we consider the correlations between the
visibilities at two different baselines $\mathbf{U}_n$ and 
 $\mathbf{U}_m$ and two different frequencies $\nu_a$ and $\nu_b$
\begin{equation}
V_2(\mathbf{U}_n,\mathbf{U}_m,\nu_a,\nu_b) = \langle
\mathcal{V}(\mathbf{U}_n,\nu_a) \mathcal{V}^*(\mathbf{U}_m,\nu_b)
\rangle  \, .
\label{eq:viscor1}
\end{equation}
The angular  brackets $\langle ... \rangle$ here denote an ensemble average
over different  realizations of the simulation. The HI contribution to the 
two-visibility correlation is directly related to the 
power spectrum $P_T(\mathbf{k})$ (Bharadwaj \& Sethi 2001;
Bharadwaj \& Ali 2005). For OWFA we have 
(Ali \& Bharadwaj 2014), 
\begin{eqnarray}
&\,&V_2(\mathbf{U}_n,\mathbf{U}_m,\nu,\nu + \Delta \nu)
=\left( \frac{\del B}{\del T}\right)^2 \int d^2U'
  \tilde{a}(\mathbf{U}_n - \mathbf{U}', \nu) \nonumber \\ 
&\times &  \tilde{a}^*(\mathbf{U}_m - \mathbf{U}',\nu + \Delta \nu) \left[
    \frac{1}{\pi r^2}   \int_0^{\infty} dk_{\parallel} {\rm cos}(k_{\parallel}
    r' \Delta   \nu) P_T( {\bf k})\right]  \,  
\label{eq:viscor2}
\end{eqnarray}
where $\tilde{a}(\mathbf{U}, \nu)$ is the Fourier transform of the beam
pattern $A(\bm{\theta}, \nu)$ and $\mathbf{k}$  refers to a comoving wave
vector with radial and transverse  components $k_{\parallel}$ and $\left( 2
\pi /r \right) \mathbf{U}'$ respectively. In evaluating this expression
it has been assumed that for all the terms in the {\it r.h.s.} 
the variation with $\nu$  is much slower in comparison with the
variation with  $\Delta \nu$. As a consequence we have  held the values of
$r$,$r^{'}$, $(\partial B/\partial T)$ and $\mathbf{U}_n$ 
fixed at that corresponding to the central frequency.

Earlier work (Ali \& Bharadwaj 2014) has shown that for OWFA the visibility
correlation $V_2(\mathbf{U}_n,\mathbf{U}_m,\nu_a,\nu_b)$ has non-zero values
only for the situation when  the two baselines are the same ($m=n$) or if 
they are adjacent  $( |m-n|= 1) $, the correlation is zero in all
other situations. Further, the correlation is approximately four times smaller
if we consider adjacent baselines instead of the
same  baseline (Bharadwaj et al. 2015). In this paper we have restricted the analysis
to the situation where $n=m$  and we have 
considered
\begin{equation}
 V_2(\mathbf{U}_n,\Delta \nu) \equiv
V_2(\mathbf{U}_n,\mathbf{U}_n,\nu_a,\nu_b)
\end{equation}
with $\Delta \nu = \mid  \nu_a -\nu_b \mid $ to validate the simulations. 

We have used $180$ independent realizations of the simulated visibilities to
evaluate the two-visibility correlation 
$V_2(\mathbf{U}_n,\Delta \nu)$ 
(eq. \ref{eq:viscor1}).  The results thus obtained were compared with the
linear theory  predictions  (eq. \ref{eq:viscor2}) in order to validate the
simulations. 

\subsection{Results}
We first consider the visibility correlation $ V_2(\mathbf{U}_n,0)$
which corresponds to the situation where $\Delta \nu=0$.  In this case  we have 
 ${\rm cos}(k_{\parallel} r' \Delta   \nu) =1$  in eq. (\ref{eq:viscor2}) for which 
the visibility correlation  $ V_2(\mathbf{U}_n,\Delta \nu)$ is maximum.

\begin{figure}[h!]
\psfrag{DECCCCCCCCCC}{$V_2\left( U_n, 0\right) \;$ Jy$^2$}
\psfrag{RA}{$U_n$}

 \centering \includegraphics[scale=0.5, angle =
  270]{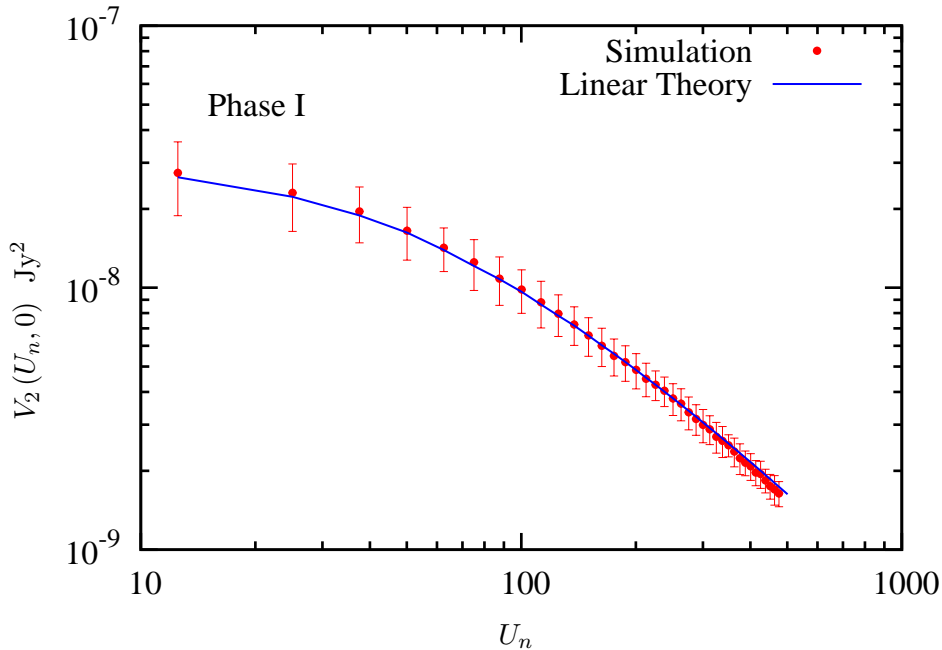}
\caption{ The  solid curves shows the linear theory predictions
  (eq. \ref{eq:viscor2})  for the  HI
  signal  $V_2\left( U, \Delta \nu \right) $  for $\Delta \nu =
  0$,  and the  points show the results from the simulations
  (eq. \ref{eq:viscor1}). The $1 \sigma$ error bars on the 
  simulated data points are estimated  
from $180$ independent realizations of the simulated visibilities.}
\label{fig:3}
\end{figure}

 Figure \ref{fig:3} shows a comparison between the linear theory predictions
 (line) and the simulations  
(points).  The $1 \sigma$ error bars on simulated data points are 
estimated from the $180$ independent realizations of the simulated
visibilities. We see that the simulations  
are in very good agreement with the  linear theory predictions. 

\begin{figure}[h!]
\psfrag{RAAAAAAA}{$\Delta \nu \; {\rm MHz}$} \psfrag{DECCCCCC}{$\kappa_{U_n}
  \left( \Delta \nu \right) $}
\psfrag{Base = 12.5} {$\;U_{1} = 12.5$} 
\psfrag{Base = 125} {$U_{10} = 125\,$} 
\psfrag{Base = 250} {$U_{20} = 250\,$} 
   \centering \includegraphics[scale=0.5,
  angle = 270]{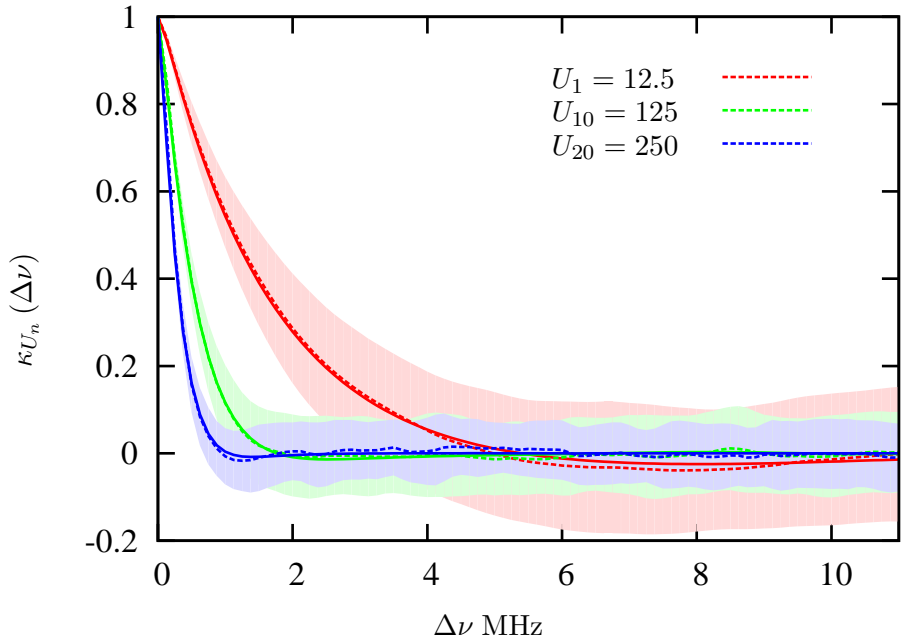}
\caption{ This  shows the  frequency decorrelation
  functions $\kappa_{U_n} \left( \Delta \nu \right) $ as a function of
  $\Delta \nu $ at three different $U_n$ values. Solid lines show the
linear theory  predictions and dashed lines are the results obtained from the
  simulations.The shaded regions show the 
$1 \, \sigma$ variation measured from $180$ realizations of the simulated
  visibilities. We see that the signal decorrelates more rapidly
at the longer baselines as compared to   the shorter ones. }  
\label{fig:4}
\end{figure}

We next consider how the visibility correlation $V_2(\mathbf{U}_n,\Delta \nu)$ varies with respect to 
$\Delta \nu $.  It is convenient to use the frequency decorrelation function
$\kappa_{U_n} \left( \Delta \nu \right) $ (Datta, Roy Choudhury \&
Bharadwaj 2007) which is defined as 
\begin{equation}
\kappa_{U_n} \left( \Delta \nu \right) = \frac {V_2\left( \mathbf{U}_n, \Delta \nu
  \right)} {V_2\left( \mathbf{U}_n, 0\right)} \,.
  \label{eq:dcor}
\end{equation}
This quantifies how rapidly  the HI signal decorrelates
as we increase the frequency separation $\Delta \nu $. The
correlation is maximum for $\Delta \nu = 0$ where $\kappa_{U_n} \left(
\Delta \nu = 0 \right) = 1$, and the linear theory  analytic calculations
(Ali \& Bharadwaj 2014)  predict the correlation to fall as $\Delta \nu $ is
increased.  Figure \ref{fig:4} shows the variation of the decorrelation
function as a  function of $\Delta \nu $ for different values of $U_n$.  
We observe that the simulations (dashed lines) are 
in good agreement with the linear theory predictions (solid lines) for all values of $U_n$. 
The HI signal decorrelates slowly with increasing $\Delta \nu$ at the short baseline
(e.g. $U_n = 12.5$ for $n = 1$)  in comparison to the long baselines 
(e.g. $U_n = 250$ for $n = 20$) where the HI signal decorrelates relatively rapidly. 
We see that for $n=1$ we have 
$\kappa_{U_n} \left( \Delta \nu \right) = 0.5$ at $\Delta \nu \approx 1 \, {\rm MHz}$
 beyond which the value of $\kappa_{U_n} \left( \Delta \nu \right)$ falls further. In 
comparison, for $n=20$ we have $\kappa_{U_n} \left( \Delta \nu \right) = 0.5$ at a smaller frequency
separation of  $\Delta \nu \approx 0.4 \, {\rm MHz}$. 
 
In summary here, we note that our results show that the simulations faithfully reproduce both 
the $U_n$ and the $\Delta \nu$ dependence of the visibility correlation 
$ V_2(\mathbf{U}_n,\Delta \nu)$ expected from the  linear theory  predictions 
(eq. \ref{eq:viscor2}) thereby validating the simulations.

\section{Particle Based Simulation}
The simulations presented in the previous section treat  the HI signal entirely as 
a diffuse radiation field and are completely oblivious to the discrete nature of the
astrophysical sources  which host the HI.  While each  discrete
source  subtends an angle that is much smaller than the  angular
resolution of  OWFA, the velocity structure of the HI inside the
individual sources  is well within reach of OWFA's frequency
resolution and this is expected  to  have an impact on the observed
HI signal.In this section we outline a particle based simulation 
which allows the flexibility of introducing any desired line profile for the 
radiation from the individual sources. The simulation technique  here follows an
 earlier  work  (Bharadwaj \& Srikant 2004).

\begin{figure}
\centering \psfrag{DDDDDECCCCCCC}{\scalebox{1.5}{ $ \quad \quad
    \frac{P_{\textrm{N-body}}(k)}{P_{\textrm{Linear}}(k)}$}}
\psfrag{RAAAAAAAAAAAA}{$\quad \quad \quad k \; {\rm Mpc}^{-1}$}
\includegraphics[scale=0.5, angle = 270]{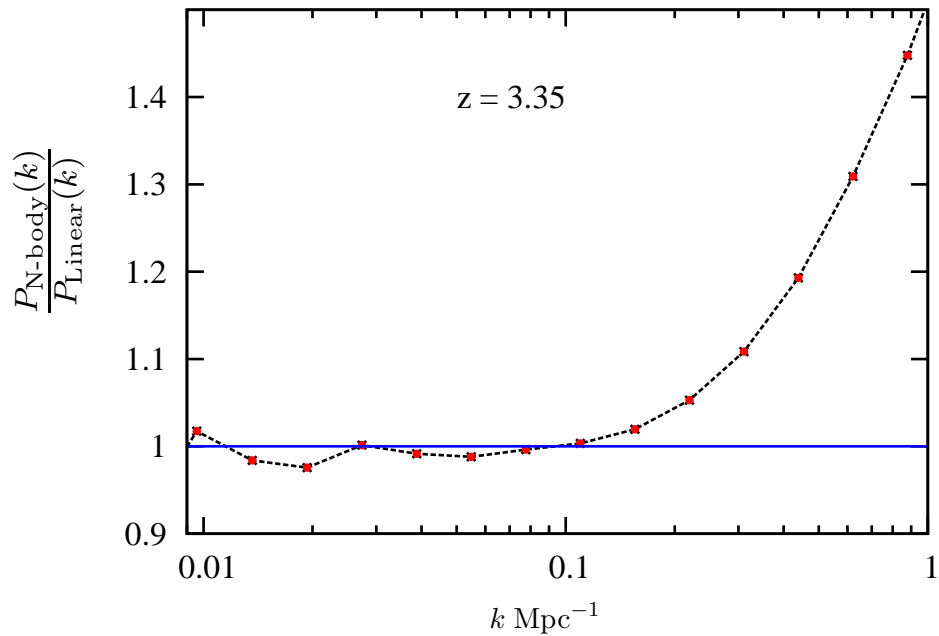}
\caption{This shows the ratio of the power spectrum of density
  fluctuations from N-body simulations and the $z=3.35$ linear theory predictions.
The power spectra shown here do not incorporate redshift space distortions.}
\label{fig:5}
\end{figure}

The simulation uses a cosmological Particle Mesh (PM) N-body code
to generate the dark matter particle distribution  at $z =3.35$. 
The simulations were done using a  $N^3 =[2048]^3$ grid which corresponds to
 a comoving volume of $[2944 \, {\rm  Mpc}]^3$ with $[1024]^3$ particles
  in it.  We have chosen the grid spacing 
$L = 1.4375 \, {\rm Mpc}$ so that it exactly matches the frequency channel width 
$L = r^{'} \times (\Delta\nu_c)$. We also incorporate the redshift space distortion effect from
the velocity information of the dark matter particles. 
Figure \ref{fig:5} shows the ratio of the power spectrum of the density
fluctuations in the dark matter distribution from N-body simulations and
the power spectrum obtained from linear theory at $z = 3.35$. We find
that the power spectrum obtained from the N-body simulation starts showing
deviations from the linear power spectrum at Fourier modes $k\geq 0.2
{\rm Mpc}^{-1}$ and the deviation increases as we move on to higher $k$
modes.  These deviations are a consequence of the
non-linear evolution of the density fluctuations at small scales. This is an additional 
feature where the present  technique improves upon the method presented 
in the previous section. We remind the reader that the previous method was 
entirely based on the predictions of linear theory.

We have cut the simulation box into 6 equal segments along the line of sight to produce 6
independent realizations each corresponding to $39 \, {\rm MHz}$ along the line of
sight. The simulation boxes were further cut to have dimensions $1130 \, {\rm Mpc}$ and 
$2944 \, {\rm Mpc}$ respectively along the $x$ and $y$  axes, and the particle distribution 
in the resulting boxes was used to calculate the HI signal. 
We assume that the hydrogen  has a mean neutral  fraction
$\bar{x}_{HI} = 2.02 \times 10^{-2}$ and the total HI content is equally distributed amongst all 
the N-body particle, whereby distribution follows the underlying particle  density $n(\x)$.
 Incorporating this, 
we model the redshifted 21-cm brightness temperature distribution in the simulation 
box  as 
\begin{equation}
T(\x)=\bar{T} \, \xb \, b_{\HI} \, \frac{n(\x)}{\bar{n}}
\label{eq:b1}
\end{equation}
where the factor $b_{HI} = 2.0$ introduces the linear bias, ${\bar{n}}$ is the mean particle density, 
and 
\begin{equation}
n(\x) = \sum_{i=1}^N \delta^3(\mathbf{x}-\mathbf{x}_i)
\label{mun_den}
\end{equation}
where $\x_i$ refers to the positions of the individual particles
in redshift space. It is important to note that the impact of the discrete nature of the HI sources 
depends on ${\bar{n}}$ the number density of HI clouds. The actual number density of the HI
cloud around $z = 3.35$ is not known, and in this work we have arbitrarily set 
${\bar{n}} = 0.04 \, {\rm Mpc}^{-3}$ which corresponds to the particle number density in the simulation.

In the last section we described how we map the brightness temperature
 distribution $T(\x)$ to a brightness temperature 
distribution $T(\bm{\theta}, \nu)$ in angle and frequency in order to calculate 
the visibilities (eq. \ref{eq:vis1}).  The position of each  simulation particle 
$\x_i$ also undergoes a similar mapping.  
As mentioned earlier, the two transverse dimensions
are mapped to  angles ($\theta_{ix}$, $\theta_{iy}$) = ($x_i/r$,  $y_i/r$)
and the radial direction is mapped to frequency 
($\nu_i= 307 \, {\rm MHz}+ z_i/r^{'}$)  where all the positions and angles
 are all with reference to the box  center.  Here, the HI radiation from 
a particular source at a radial distance of $z_i$ will be received at a
single frequency  $\nu_i$ which is determined by the cosmology, and the particular 
source will not contribute to any
of the other frequencies. In reality, the HI emission emanating from any astrophysical 
source will have a spread  in frequencies due to a variety of reasons including the 
thermal  line broadening and the bulk motions of the HI gas inside the source.  We include this 
in the simulation by introducing a line profile $\phi_i(\nu-\nu_i)$ associated with each source. 
The line profile determines the frequency spread of the redshifted HI 21-cm emission from a 
particular source, and  it is normalized as 
\begin{equation}
\int \phi_i(\nu) d\nu = 1
\label{line_prof}
\end{equation}
While the line profile may, in principle, differ from source to source 
we have here assumed the same line profile for all the sources in order to keep 
the analysis simple. It may further be noted that the line profile here has been defined with respect to 
the observed frequency in the rest frame of the telescope.  Using eq (\ref{eq:b1}) in eq. (\ref{eq:vis1})
 and incorporating the line profile, we obtain a relation which allows us to calculate 
the OWFA visibilities corresponding to 
the simulated  particle distribution 
\begin{equation}
\V(U_n,\nu_a)=  
\left( \frac{\del B}{\del T}\right) \frac{\bar{T}}{\bar{n} r^2 r'} 
 \sum_i A(\bm{\theta}_i, \nu_i) \phi(\nu_a -\nu_i) 
e^{2\pi i U_n \theta_{yi}} \,.
\label{eq:b2}
\end{equation}
While we have the freedom of implementing any desired line profile
for the HI sources, for the purpose of this paper we have chosen 
a very simple line profile 
\begin{eqnarray}
\phi(\nu-\nu_i) &=& \frac{1}{F\,\Delta \nu_c} \, \hspace{1cm} {\rm for} \, 
\mid \nu - \nu_i \mid  \le \frac{F}{2} \times \Delta \nu_c 
\label{eq:b3}\\
\, &=& 0 \,\hspace{3cm} {\rm otherwise} \,. \nonumber 
\end{eqnarray}
Here we have considered three values of $F = 1,\,2,\,3$. For $F=1$
every source here is 
assigned a flat line profile of width $0.125 \, {\rm MHz}$, which is also the frequency resolution
assumed for OWFA,  this corresponds to a velocity spread of $114 \, {\rm km s}^{-1}$ at the rest frame 
of the source, which is a reasonable value for the HI velocities inside the individual 
sources expected at this redshift. For $F=2 \,{\rm and} \, 3$ the assigned flat line profiles are of broader width $0.25 \, {\rm MHz}$ and $0.375 \, {\rm MHz}$, which respectively correspond to twice and thrice the frequency resolution of OWFA.
\subsection{Results}
\begin{figure}
\centering \psfrag{DECCCCCCCCCC}{$V_2\left( U_n, 0\right) \;$ Jy$^2$}
\psfrag{RA}{$U_n$}
 \includegraphics[scale=0.5, angle = 270]{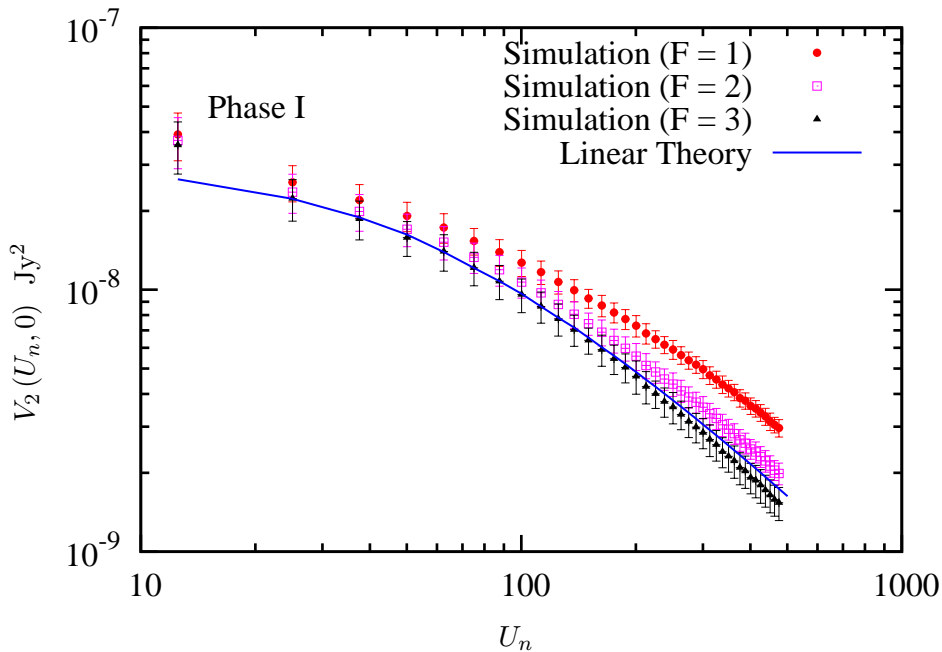}
\caption{The solid curve shows the linear theory predictions, and the points
  in different styles (see legend) show the results from the  
 simulations. The $1 \,\sigma$ error bars are  estimated from  $180$
 realizations of the simulations.}
\label{fig:6}
\end{figure}

Figure \ref{fig:6} shows the visibility correlation $V_2\left( U_n,
0\right)$ determined from the simulations. We have run $180$ independent
realizations of the simulation for   which the mean  and  the standard
deviation are respectively shown with points and the corresponding error
bars.  We find that for all values of $F$ 
the value of $V_2\left( U_n,0\right)$ obtained from the simulations is
approximately twice the linear 
prediction for the shortest baseline $U_n$ with $n=1$. For $F=1$ 
the values obtained from the simulations are considerably 
larger than the    linear theory prediction (solid line) for all values of
$U_n$. 
The simulated values are  roughly a factor of $1.3$ times larger than the
linear prediction in the baseline  
range $20 \leq U_n \leq 100$ which corresponds to $ 2 \le n \le 8$ , the
deviation between the simulations and the linear theory predictions 
is found to increase at longer baselines $U_n > 100$. Barring the shortest
baseline, the values  
of $V_2\left( U_n, 0\right)$  progressively decrease as  $F$  is increased to 2 and 3.
For $F = 2$ the results are still in excess of the linear predictions. For $F
=3$ the results match with the  linear theory predictions for $U_n < 100$
whereas the results are below the linear theory predictions for  larger
baselines. The deviations from linear theory seen for $F=1,2$ and $3$
can be attributed   to a combination of the effect of the finite line width 
(eq. \ref{eq:b3}) and  the non-linear evolution of the particle distribution in
the N-body simulations (Figure \ref{fig:5}).  However, we expect the
non-linear evolution to be more significant at the longer baselines, and 
it is difficult  to  use these two effects to explain  why  the shortest 
baseline shows a larger  deviation compared  to the other  adjacent longer
baselines for all values of $F$.  

The earlier work (Ali \& Bharadwaj 2014) has shown that the OWFA visibilities
only respond to the brightness temperature fluctuations $\delta
T(\bm{\theta},\nu)$, and are insensitive to the mean component
$\bar{T}_b(\nu)$. This however had assumed the flat sky approximation, and the
sky was assumed to have an infinite extent. This however does not hold for the
simulations which also assume the flat sky approximation but have a finite sky
extent.  The ${\rm sinc}^2$ primary beam pattern (eq. \ref{eq:pbeam})
is now truncated after a finite extent and
as a consequence the visibilities all pick up a contribution from
$\bar{T}_b(\nu)$. This contribution could be relatively large for the short
baselines, however it is expected to die down at the longer baselines. We can
explain the large deviation at the shortest baseline by attributing it to the
mean component $\bar{T}_b(\nu)$  that is picked up due to the finite sky
coverage. We note that this problem does not arise in the previous section
where the mean component is not included in the simulations.

\begin{figure}
  \psfrag{ratioratioratio}{$\quad \frac{V_2(U_n,0)}{V_2(U_1,0)}$}
  \psfrag{Un}{$U_n$}
  \centering \includegraphics[scale=0.5,
 angle = 270]{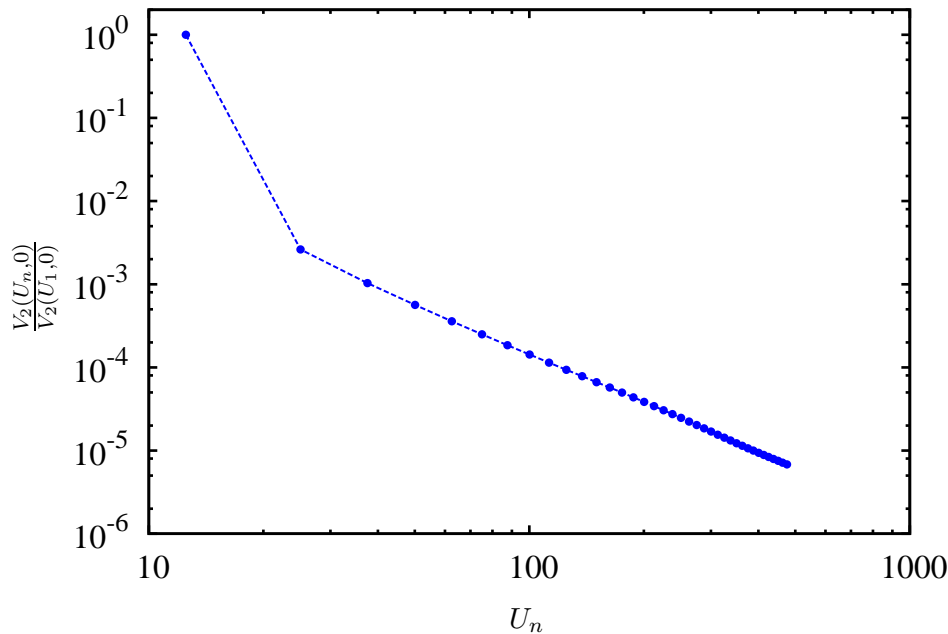}
\caption{ Considering a simulation where the brightness
temperature distribution has only a mean contribution 
 $\bar{T}_b(\nu)$, this shows the ratio $V_2(U_n,0)/V_2(U_1,0)$
as a function of $U_n$.}
\label{fig:7a}
\end{figure}

In order to verify our understanding of  the origin of the large deviation at
the shortest baseline, we have considered a situation where the brightness
temperature distribution (eq. \ref{eq:fluc})  has only a contribution from 
 $\bar{T}_b(\nu)$ and  the  fluctuations  $\delta T(\bm{\theta},\nu)$
have been set to zero. These simulations were run for $F = 1$ only. Considering the ratio  $V_2(U_n,0)/V_2(U_1,0)$
(Figure \ref{fig:7a}) we find that  this falls by 
approximately three orders of magnitude as we go from $n=1$ to $n=2$. The
ratio falls even further if we consider longer baselines $n > 2$.
This supports our interpretation that
in our simulations the shortest  baseline picks up a significant contribution
from  the mean brightness temperature, however this contamination is not
significant at the longer baselines.


\begin{figure}
\psfrag{RAAAAAAA}{$\Delta \nu \; {\rm MHz}$} \psfrag{DECCCCCC}{$\kappa_{U_n}
  \left( \Delta \nu \right) $}
\psfrag{Base = 12.5} {$\;U_{1} = 12.5$} 
\psfrag{Base = 125} {$U_{10} = 125\,$} 
\psfrag{Base = 250} {$U_{20} = 250\,$}   
   \centering \includegraphics[scale=0.5,
  angle = 270]{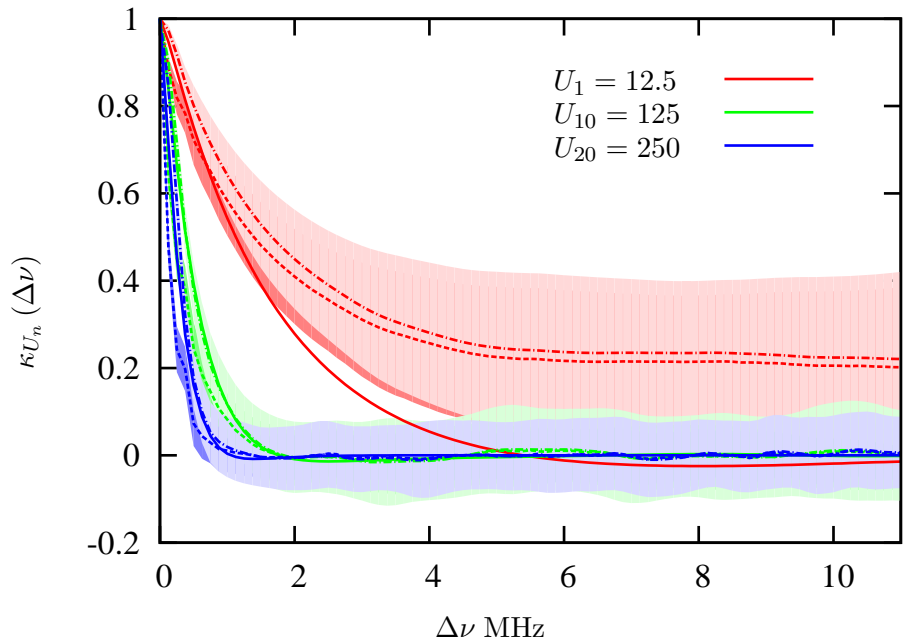}
\caption{ This shows the linear theory  (solid line) and  simulated  
(dashed line for $F=1$, dot-dashed line for $F=3$) 
 $\kappa_{U_n} \left( \Delta \nu \right) $ as a function of
  $\Delta \nu $ at three different $U_n$ values.}
\label{fig:7}
\end{figure}
Figure \ref{fig:7} shows the variation of $\kappa_{U_n}(\Delta \nu)$ with
$\Delta \nu$ for three different baselines. Notice that 
 $\kappa_{U_n}(\Delta \nu)$ obtained from the simulations at the shortest
baseline  saturates at a
relatively large value $(\sim 0.25)$ at  frequency separations 
$\Delta \nu > 4 \, {\rm MHz}$ in contrast to the linear predictions which are
close to zero in this range. A similar discrepancy 
 is not seen at the longer baselines where
both the simulations and the linear prediction are close to zero for large
$\Delta \nu$. This discrepancy at the shortest baseline also can be explained
by the fact that the shortest baseline picks up a relatively large
contribution from $\bar{T}_b(\nu)$ due to the limited angular extent of the
simulations. This problem does not seem to be significant at the longer
baselines, and we have dropped  the shortest baseline from the subsequent
discussion of this section.   
\begin{figure}[h!]
\psfrag{DECCCCCCCCCC}{$V_2\left( U_n, \Delta \nu\right) \; {\rm Jy}^2$}
\psfrag{RAAAAAAA}{$\Delta \nu \,{\rm MHz}$}
\psfrag{n = 12} {$U_{12} = 150$}
\centering \includegraphics[scale=0.5, angle =270]{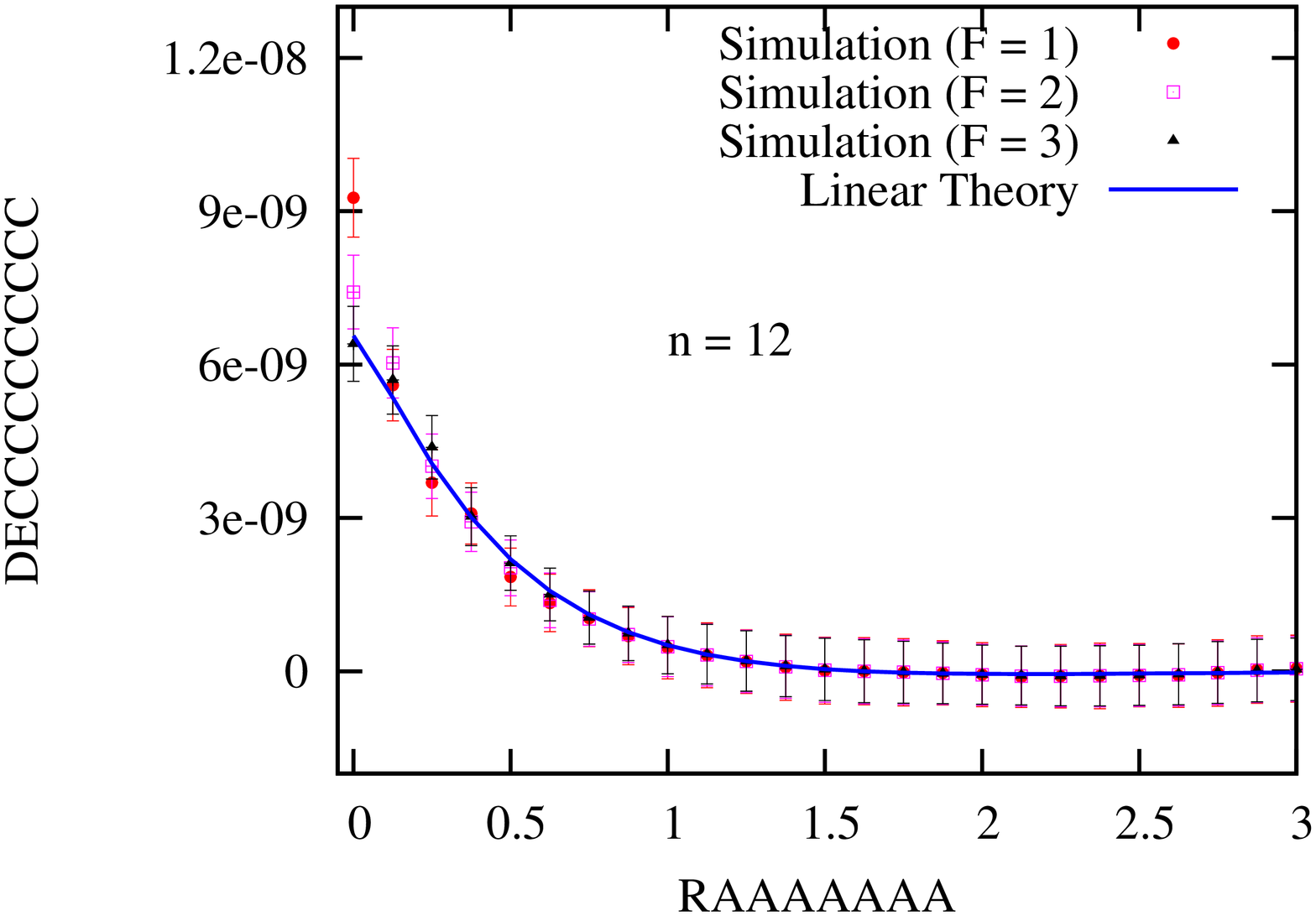}
\caption{ The solid curve shows the linear theory predictions for the visibility correlation 
$V_2\left( U_n, \Delta \nu \right) $, and the points in different styles (see legend) show the results from the 
 simulations as a function of $\Delta \nu$ for
  $U_n=150$ with  $n=12$.}
\label{fig:a}
\end{figure}
\begin{figure}[h!]
\psfrag{DECCCCCCCCCC}{$V_2\left( U_n, \Delta \nu\right) \; {\rm Jy}^2$}
\psfrag{RAAAAAAA}{$\Delta \nu \, {\rm MHz}$}
\psfrag{n = 36} {$U_{36} = 450$}
\centering \includegraphics[scale=0.5, angle =270]{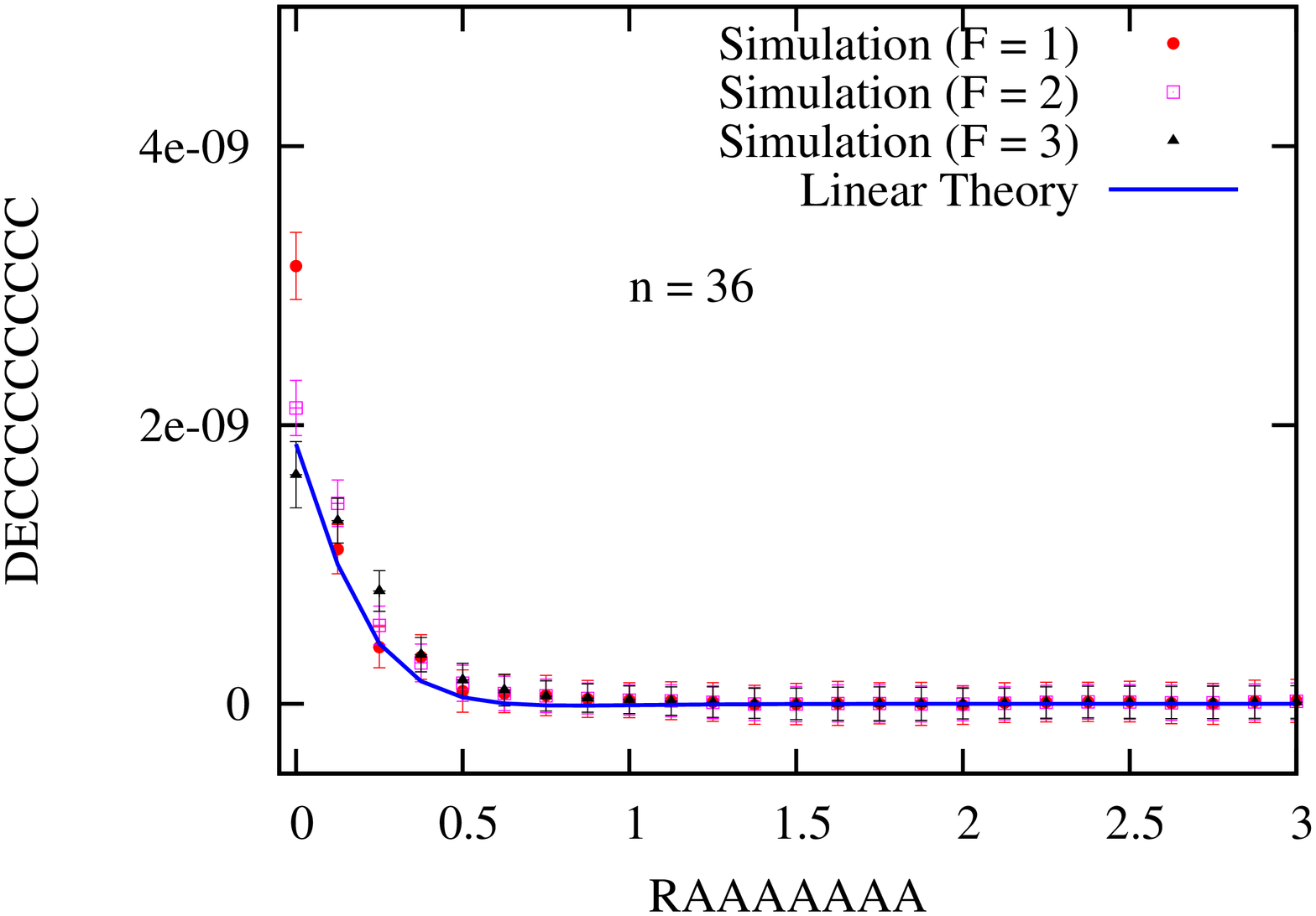}
\caption{ The solid curve shows the linear theory predictions for the visibility correlation 
$V_2\left( U_n, \Delta \nu \right) $, and the points in different styles (see legend) show the results from the 
 simulations as a function of $\Delta \nu$ for
  $U_n=450$ with  $n=36$.}
\label{fig:b}
\end{figure}

For these simulations we find that it is more convenient to directly study
$V_2(U_n,\Delta \nu)$ instead of considering  $\kappa_{U_n}(\Delta \nu)$ 
to  analyze how the visibility correlation behaves 
 as a  function of $\Delta \nu $.  Figures \ref{fig:a} and \ref{fig:b} 
show  $V_2(U_n,\Delta \nu)$  as a  function of $\Delta \nu $ 
 for the two  values  $U_n=150$  and $450$ respectively.  We have already seen
 that for $F=1$ the values of $V_2(U_n,0)$ from the simulations are $\sim 1.3$
 times  larger than the linear prediction in the range $20 \le U_n \le 100$ 
 (Figure \ref{fig:6}), and this discrepancy increases at longer baselines 
$U_n > 100$. Here we find that for $F = 1$, $V_2(U_n,0)$ from the simulations are 
$\sim 1.5$ times larger than the linear prediction at $U_n=150$
(Figures \ref{fig:a}). The  discrepancy in $V_2(U_n,0)$ 
is reduced for $F = 2$ and the results are  
roughly consistent with linear theory for $F =3$.
We further see that the discrepancies are  restricted to 
$\Delta \nu=0$  only for all values of $F$. The simulations are consistent
with the linear  predictions for larger values of $\Delta \nu$,  except
possibly for a slight excess 
 seen at $\Delta \nu=0.125 {\rm MHz}$ for $F = 2$ . The discrepancy at
 $\Delta \nu=0$ arises from the line  profile  $\phi(\nu)$. The HI
 emission from any individual source in the simulation will be   
spread across a finite frequency width introducing an excess correlation within
the line width (Bharadwaj \& Srikant 2004).  
For $F=1$ the line width is the same as the channel width
 $\Delta \nu_c$ which is why the discrepancy is only limited to $\Delta \nu=0$
and does not extend to larger frequency separations for $F = 1$. For $F=2$ the
line width is doubled and its  height is halved. This explains why the
deviation from linear theory in $V_2(U_n, 0)$  drops relative to
$F=1$. This is also possibly responsible for the small excess that we see 
at $\Delta \nu = 0.25 \, {\rm MHz}$ for $F=2$. For $F=3$ the
line width is thrice that for $F=1$ whereas its  height is three times
smaller. There is no noticeable excess correlation seen for $F=3$ because of
the reduced line height, and the 
results are consistent with linear theory for all values of $\Delta \nu$. 

We next consider a larger baseline $U_n=450$ for which $V_2(U_n,\Delta \nu)$
is  shown in  Figure \ref{fig:b}.   For $F = 1$ we find that the value 
obtained from the  simulations is $\sim 1.5$ times larger than the linear
prediction for $\Delta \nu=0$ and the simulations are consistent with the
linear predictions for all larger values of $\Delta \nu$. For $F=2$ the
discrepancy  from the linear predictions reduces in magnitude but it now
extends to  $\Delta = \nu=0$ and  $0.125 \, {\rm MHz}$. For $F=3$ the result
from the simulation falls below the linear prediction at  $\Delta \nu=0$ while
we have an excess at $\Delta \nu0.125 \, {\rm MHz}$ and $0.25 \, {\rm MHz}$. 
In summary we find that  the magnitude and frequency extent of the deviations
from the linear predictions  is strongly correlated with the value of $F$
which decides the frequency width and height of the \HI emission line from the
individual sources. This leads to a picture where the deviations from linear
theory seen in Figures  \ref{fig:6}, \ref{fig:7},\ref{fig:a} and \ref{fig:b}
can be largely attributed to the effect of the line profile of the individual
\HI sources. In the present analysis we are unable to identify any noticeable 
signature  of he non-linear evolution of the underlying dark matter
distribution.  
Finally we note that the good agreement between the
simulations and the linear predictions found at the long
baselines for large frequency separations confirms the fact that the longer  
baselines do not pick up a significant contribution from $\bar{T}_b(\nu)$. 

\section{Discussion}
In this paper we present two different methods for simulating the
redshifted HI 21-cm visibility signal  expected in observations with Phase I of
OWFA. The first method directly generates random realizations of
the  brightness temperature fluctuations $\delta T(\bm{\theta}, \nu)$
corresponding to the input model  power spectrum $P_T(\mathbf{k})$ at $z =
3.35$ (eq. \ref{eq:temp_pow}). These simulations treat the HI signal
entirely as a diffuse radiation and ignores the fact that the 
HI actually resides in  discrete clouds. Further, it also assumes that the
redshifted 21-cm radiation is a Gaussian random field whose properties are
completely described by the power spectrum $P_T(\mathbf{k})$ calculated using 
linear perturbation theory. The second method uses a
cosmological  N-body simulation to generate a  $z=3.35$
particle distribution. We identify these particles as the discrete sources
that host the HI. This method allows us the freedom to assign any desired line
profile to the 
HI emission from the individual sources.  It also captures the non-linear
evolution of the density field. For the present work we have used a simple
line profile where  each source is assigned an uniform  line width. The simulations
were done for three different line widths 
$\Delta \nu = 0.125 \, {\rm MHz}, 0.25 \, {\rm MHz} \,{\rm and}\, 0.375 \,
{\rm MHz}$
which respectively  correspond to $F=1, 2 \,{\rm  and}\, 3$ in our model for
the line profile.

The simulated visibilities, in both cases, are a random signal. We have
considered the two-visibility correlation $V_2(U_n,\Delta \nu)$ to analyze the
results of  the simulations and compare them with the predictions of linear
theory. As expected, we find that the first method faithfully reproduces
the predictions of linear theory. For the second method, we find that the
visibilities at the shortest baseline pick up a relatively large contribution 
from $\bar{T}_b(\nu)$ possibly due to the limited sky coverage of the simulations. The
visibilities at the longer baselines are not significantly contaminated by a
contribution from 
$\bar{T}_b(\nu)$. We consider two long baselines, $U_n =
150 \, {\rm and} \, 450$ respectively in order to analyze the results of the
second method. We find that the results from the simulations show
discrepancies from the linear theory predictions. These discrepancies are
restricted to $\Delta \nu$ values which are comparable to or smaller than  
the frequency width of the line profile, and the simulations are
consistent with linear theory at larger $\Delta \nu$. 
The magnitude of the discrepancy depends on the width of the line profile, and
for $F=1$ we find an excess over the linear predictions at  $\Delta\nu=0$. 
The magnitude of this excess  comes down  when $F$ is changed from $1$ to
$2$. We interpret this as arising from the fact that height of the line
profile is halved when $F$ is changed from $1$ to $2$.  We find that the
simulations fall below the linear theory prediction at  $\Delta \nu=0$
for $F=3$. In this case the increased line width  possibly smears out some of
the correlation, and  the line height is also further reduced .

In conclusion we note that simulated data generated using the methods
presented in this paper will play an important role in validating the OWFA
data analysis pipeline. Simulated data will also play an important role in
correctly interpreting the observed \HI signal once a detection is made. 
In this work we find that it is important to take into account  the discrete 
nature of the \HI coulds. The line profile of the individual \HI clouds play
an imprtant role in shaping the statistical properties of the redshifted 21-cm 
signal at small frequency separations which are comparable to the line width
of the emission from the individual clouds.   We note that similar 
results have been reported in an earlier paper (Bharadwaj \& Srikant
2004). For the present work we have used a rather simplistic line
profile, we plan to incorporate more
realistic models for the \HI line profile in future work.
 Further, the impact of the discrete nature of the HI sources 
depends on ${\bar{n}}$ the number density of \HI clouds which has been set
arbitrarily  to ${\bar{n}} = 0.04 \, {\rm Mpc}^{-3}$ which corresponds to the
particle  number density in the simulation. We plan to consider more realistic \HI
 mass distribution schemes in future work. 

The current work  reveals that the flat-sky paradigm 
adopted here fails to correctly model the visibilities at the shortest
baseline, and it is necessary to implement the  spherical sky in order
to correctly handle this. This is also expected to be important for OWFA
Phase II which will have a much larger field of view. We plan to also address
this  in subsequent work. 
\section*{Acknowledgement}
The authors would like to thank Jayaram Chengalur for the useful suggestions
and discussions.  SC acknowledges the University Grant Commission, India for
providing  financial support through Senior Research Fellowship. SC would also
like to thank Anjan Kumar  Sarkar and Debanjan Sarkar for their help and
useful discussions. 

\end{document}